\documentclass[a4paper,10pt]{article}

\usepackage[utf8x]{inputenc}
\usepackage[T1]{fontenc}

\usepackage[pdftex]{graphicx} 

\usepackage[a4paper, lmargin=0.1666\paperwidth, rmargin=0.1666\paperwidth, tmargin=0.1111\paperheight, bmargin=0.1111\paperheight]{geometry} 

\usepackage{natbib}

\usepackage[version=3]{mhchem} 
\usepackage{graphicx}
\usepackage{siunitx}
\usepackage{xcolor}
\usepackage{color}
\usepackage{listings}
\usepackage{amsmath,amssymb}
\usepackage{ifthen}

\definecolor{backcolour}{rgb}{0.95,0.95,0.92}
\lstdefinestyle{mystyle}{
    backgroundcolor=\color{backcolour},
    basicstyle=\ttfamily\footnotesize,
    frame=none
}
\newcommand*\lsin{\lstinline[basicstyle=\ttfamily\footnotesize]}

\newboolean{bShowFigs}
\setboolean{bShowFigs}{true}

\begin{document}

\subsubsection*{Title}
{\Large
Predicting solution scattering patterns with explicit-solvent molecular simulations}

\subsubsection*{Running title}
Explicit-solvent SAS predictions

\subsubsection*{Authors}
Leonie Chatzimagas$^1$ and Jochen S.\ Hub$^1$

\subsubsection*{Affiliations}
$^1$Theoretical Physics and Center for Biophysics, Saarland University, Saarbrücken 66123, Germany

\subsubsection*{Keywords}
Small-angle scattering, SAXS, SANS, all-atom molecular dynamics simulations, hydration layer, excluded solvent




\subsubsection*{Abstract}
Small-angle X-ray or neutron scattering (SAXS/SANS/SAS) is widely used to obtain structural information on biomolecules or soft-matter complexes in solution. Deriving a molecular interpretation of the scattering signals requires methods for predicting SAS patterns from a given atomistic structural model. Such SAS predictions are non-trivial because the patterns are influenced by the hydration layer of the solute, the excluded solvent, and by thermal fluctuations. Many computationally efficient methods use simplified, implicit models
for the hydration layer and excluded solvent, leading to some uncertainties and to free parameters that require fitting against experimental data. SAS predictions based on explicit-solvent molecular dynamics (MD) simulations overcome such limitations at the price of an increased computational cost. To rationalize the need for explicit-solvent methods, we first review the approximations underlying implicit-solvent methods. Next, we describe the theory behind explicit-solvent SAS predictions that are easily accessible via the WAXSiS
web server. We present the workflow for computing SAS pattern from a given molecular dynamics trajectory. The calculations are available via a modified version of the GROMACS simulations software, coined GROMACS-SWAXS, which implements the WAXSiS method. Practical considerations for running routine explicit-solvent SAS predictions are discussed.

\section{Introduction}

Small- and wide-angle X-ray scattering (SWAXS) has originally been used to probe the overall shape of biomolecules or soft-matter complexes in solution. In recent decades, SWAXS has developed into an increasingly quantitative probe, primarily due to improved sample preparation, more brilliant light sources, and single-photon counting detectors \citep{koch_small-angle_2003,putnam_x-ray_2007,graewert_impact_2013}. Thanks to the advent of free-electron lasers, time-resolved SWAXS (TR-SWAXS) measurements are capable of tracking ultrafast dynamics of  biomolecules or small molecules in solution \citep{arnlund_visualizing_2014,levantino_ultrafast_2015,brinkmann_ultrafast_2016}. Complementary, small-angle neutron scattering (SANS) has remained popular because it allows for contrast-variation experiments by changing the \ce{D2O} concentration of the buffer \citep{gabel2015}. Henceforth, we use the term SAS (small-angle scattering) when referring to both SWAXS and SANS.

Analysis of the SAS curve gives information on the overall shape and macroscopic properties of the probe, such as the radius of gyration, maximum particle size,
the structural order (globular vs.\ unfolded), molecular mass, and particle volume.
In contrast, an atomistic interpretation cannot be drawn from the data alone owing to the low information content of the SAS curves, but instead requires methods for theoretically predicting SAS intensities from a given atomistic structure. Such SAS curve predictions, so-called forward models, enable the validation or the refinement of structural models or ensembles against SAS data.

The physical properties that give rise to a solution scattering curve $I(q)$ are well understood. Namely, the intensities are given by
\begin{equation}
	I(q) = \langle |F(\mathbf{q})|^2 \rangle_\Omega \, ,
\end{equation}
where $\langle \cdot \rangle_\Omega$ denotes the orientational average in reciprocal space, and $F(\mathbf{q})$ is the Fourier transform of the electron density contrast $\Delta\rho(\mathbf{r})$ between the solution and the pure-buffer system:
\begin{equation}
	F(\mathbf{q}) = \int \Delta\rho(\mathbf{r})\,e^{-i\mathbf{q}\cdot\mathbf{r}}\,\mathrm{d}^3r
\end{equation}
However, computing $I(q)$ from a given structural model such as a crystal structure or a molecular dynamics (MD) simulation of a biomolecule is complicated by several aspects:
\begin{itemize}
	\item[(i)] Since SAS detects the electron density contrast in solution, computing $I(q)$ requires knowledge of the volume of solvent that is displaced by the solute. For a biomolecule with internal disorder and a rough surface, the displaced volume is far from obvious. In addition, as discussed below, the volume taken by atoms of a certain chemical element depends on the chemical environment, suggesting that tabulated atomic volumes are subject to marked uncertainties.
	\item[(ii)] The density of the hydration layer of biomolecules differs from the density of bulk solvent, thus contributing to the density contrast $\Delta\rho$. The hydration layer is influenced by properties such as the biomolecule's charge and geometry, the amino acid composition of the surface (anionic, cationic, hydrophobic, hydrophilic neutral), or the salt type and concentration of the buffer.
	For instance, highly charged biomolecules such as DNA/RNA were shown to exhibit a tight hydration layer \citep{pollack2011}. For biomolecules with large surface-to-volume ratio, such as intrinsically disordered proteins, the hydration layer is expected to strongly contribute to the SAS signal.
	However, how such properties determine the hydration layer is not yet understood on a quantitative level \citep{kim_saxs/sans_2016}.
	\item[(iii)] SAS curves are influenced by thermal fluctuations of the biomolecule. Using analytic models, \citet{moore_effects_2014} showed that including Debye-Waller factors or correlated atomic fluctuations influences the calculated SAS curves at relatively small scattering angles of $q \gtrsim 0.2$\,\AA{}$^{-1}$. These findings are in line with
	earlier experimental observations \citep{tiede_protein_2002} and with
	MD simulations, which revealed alterations of SAXS curves at $q \gtrsim 0.25$\,\AA{}$^{-1}$ upon modulating atomic fluctuations during the simulations \citep{chen_validating_2014}.
\end{itemize}

In the last three decades, a wide range of methods have been developed for predicting SAS intensities from atomistic structures, reflecting the increasing importance of SAS experiments for biomolecular research and the need for a physically founded interpretation of the data \citep{svergun_crysol_1995, merzel_is_2002,merzel_sassim:_2002, tjioe2007,bardhan_it_2009,ravikumar_fast-saxs-pro:_2013,yang_rapid_2009,schneidman-duhovny_foxs:_2010,schneidman-duhovny_accurate_2013,stovgaard_calculation_2010,poitevin_aquasaxs:_2011,liu_computation_2012,nguyen_accurate_2014,putnam_bcl::saxs:_2015,grudinin_pepsi-saxs:_2017,oroguchi_intrinsic_2009,oroguchi_mdsaxs_2012,grishaev_improved_2010,virtanen_modeling_2011,park_simulated_2009,kofinger_atomic-resolution_2013,chen_validating_2014,knight_waxsis:_2015}. The methods mainly differ by
\begin{itemize}
	\item the method for modeling the hydration layer, involving explicit all-atom models, increased atomic form factors for solvent-exposed solute atoms, layers of uniform electron density, implicit models with internal structure, or several others;
	\item modeling of the excluded solvent based on tabulated atomic volumes up to explicit all-atom models;
	\item mathematical methods for computing the orientational average involving the Debye equation, a spherical harmonics expansion, or numerical quadrature;
	\item the spatial resolution, i.e., whether the SAS curves are computed from individual atoms or from coarse-grained beads.
\end{itemize}

In this chapter, we focus on the calculation of SAS curves from atomistic explicit-solvent MD simulations as implemented in GROMACS-SWAXS, an extension of the GROMACS simulation software (https://gitlab.com/cbjh/gromacs-swaxs) \citep{chen_validating_2014,chen_combined_2019}. The method is easily accessible via the WAXSiS web server at https://waxsis.uni-saarland.de \citep{knight_waxsis:_2015}.
As a rationale behind the explicit-solvent methods, we first review the approximations and limitations underlying widely used implicit-solvent methods, with a focus on the role of the hydration layer and excluded solvent for obtaining the density contrast. Next, we briefly describe the theory of explicit-solvent SAS curve predictions from explicit-solvent MD simulations as used by GROMACS-SWAXS and WAXSiS. Finally, the workflow of SAS calculations from an MD trajectory and practical considerations are discussed.

\begin{figure}[h]
	\ifthenelse{\boolean{bShowFigs}}{
		\includegraphics[width=0.6\textwidth]{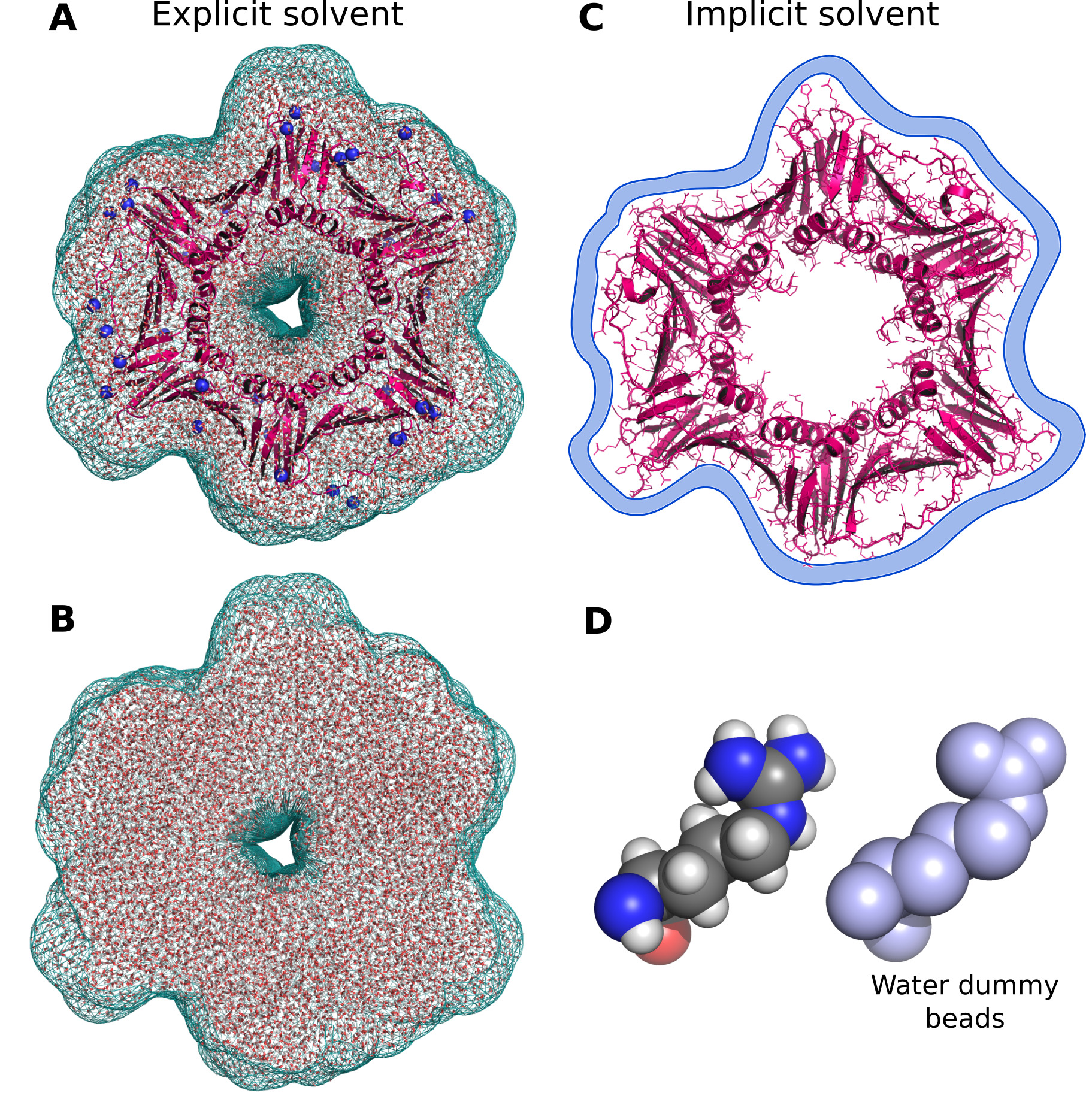}}{}
	\caption{(A/C) Hydration layer (B/D) displaced solvent, as represented by (A/B) explicit-solvent and (C/D) several implicit-solvent methods. (A) A spatial envelope is defined (blue mesh) that includes the biomolecule and the hydration layer, here shown for the protein PCNA (pdb code 4D2G) \citep{de_biasio_structure_2015}. (B) The same spatial envelope is used to define the displaced solvent. (C) Implicit-solvent methods use a simplified description of the hydration layer, often modeled as a layer of uniform density, while (D) the displaced solvent is typically described by water dummy beads with estimated volumes taken from a publication by \citet{traube_ueber_1895} and fitted against the data. \label{explsolvent}}
\end{figure}

\section{Implicit-solvent methods}

Computationally efficient methods for computing SAS curves, as implemented in CRYSOL, FoXS, PepsiSAXS, or several other tools, use an implicit representation of the hydration layer and excluded solvent. To rationalize the benefits of explicit-solvent SAS calculations, we discuss the limitation of implicit-solvent methods in the following.

\paragraph{Displaced solvent.}
Many methods represent the displaced (or excluded) solvent with the help of water dummy beads, which are typically placed at the positions of the biomolecule's heavy atoms (see Fig. \ref{explsolvent}D) \citep{svergun_crysol_1995,grudinin_pepsi-saxs:_2017}. The electron density of the bead is essentially taken as $\rho_d(r) = \rho_s \exp(-\pi r^2/v^{2/3})$, where $\rho_s$ is the electron density of the solvent and $v$ the displaced volume, such that the bead carries $\rho_s v$ electrons, as desired. The atomic form factor of the bead is given by the Fourier transform of the electron density, $f_d(q) = \rho_s v \exp(-v^{2/3}q^2/4\pi)$. In the implicit-solvent description, the buffer subtraction is carried out by reducing the atomic form factor of the biomolecule's heavy atom with the form factor of the dummy bead $f_d(q)$, leading to so-called ``reduced form factors''.

This method requires knowledge of the displaced volumes $v$ for all atoms.
However, estimates for the displaced volume differ substantially within the literature, leading to some uncertainty in the predicted SAS curve \citep{hub_interpreting_2018}.
Most SAXS prediction methods use and cite the displaced volumes reported by \citet{fraser_improved_1978}. These data originally date back to a publication in 1895 by Isidor Traube, which was based on the volume change of water in response to the solvation of various organic compounds \citep{traube_ueber_1895}. Table \ref{volumes} presents the atomic volumes reported by Fraser \textit{et al.} in \si{\angstrom}$^3$ as well as the original values by Traube in ccm/mol for several atoms or chemical groups. Traube and Fraser values are identical since $\mathrm{ccm/mol} = 1.66$\,\AA$^3$.

More recently, atomic volumes have been obtained by Pontius \textit{et al.} using Voronoi tessellation applied to the core region of high-resolution protein structures \citep{pontius_deviations_1996}. The volumes by Traube and Fraser greatly differ from the volumes obtained from crystal structures (Table \ref{volumes}, columns 2 and 3). Compared to the Pontius values, the Fraser values greatly overestimate the volumes of \ce{CH} and \ce{CH2} groups, and they underestimate the volumes of N and O atoms and of \ce{NH} or \ce{OH} groups.
Notably, Traube reported volumes of N or O atoms for alternative chemical environments such as nitro or (today outdated) ``pentavalent'' environments as well as for hydroxyl groups; however, these were not cited by Fraser, which may explain why they are not used in SAS calculations today (Table \ref{volumes}, column 4).

\begin{table}[tb]
  \caption{Displaced volumes from Voronoi tessellation of crystal structure cores by \citet{pontius_deviations_1996}, as well as from densitometric data by \citet{traube_ueber_1895}, as cited by \citet{fraser_improved_1978}. Pontius values represent averages and standard deviations from all amino acids. Traube and Fraser values are identical, using $\mathrm{ccm/mol} = 1.66\,$\AA$^3$. Fraser volumes of chemical groups represent sums of single-atom volumes.}
  \label{volumes}
  \begin{tabular}{cccc}
    \hline
    Atomic group &  Pontius \textit{et al.} [\si{\angstrom}$^3$] & Fraser \textit{et al.}, 1978 [\si{\angstrom}$^3$] & Traube, 1895 [ccm/mol] \\
    \ce{H}   &           & 5.15  & 3.1 \\
    \ce{C}   &           & 16.44 & 9.9 \\
    \ce{N}   & 8.8 (0.8)  & 2.49  & 1.5 (trivalent) \\
                         &   &   & 10.7 (pentavalent) \\
                         &   &   & 8.5--10.7 (in nitro compound) \\
    \ce{O}   & 22.3 (0.4) & 9.13  &  5.5 (carbonyl oxygen) \\
                           &  &  &  2.3 or 0.4 (hydroxy oxygen) \\
    \ce{OH}  & 23.9 (0.9) & 14.28 &  \\
    \ce{NH}  & 14.1 (0.3) & 7.64  &  \\
    \ce{CH}  & 11.8 (0.6) & 21.59 &  \\
    \ce{CH2} & 20.9 (1.8) & 26.74 &  \\
    \ce{CH3} & 33.9 (1.2) & 31.89 &  \\
    \hline
  \end{tabular}
\end{table}

Besides the uncertainty owing to the choice of tabulated atomic volumes, additional uncertainty may arise because the atomic volumes depend on the chemical environment. Whereas the volumes by Pontius \textit{et al.} represent the atomic packing in the core of stable proteins, it seems unlikely that the volumes also hold for more flexible protein surfaces, for detergent micelles, or for nucleotides. For n-hexadecane, for instance, the Pointius and Fraser volumes would imply a molecular volume of 360.4\,\AA$^3$ and 438.1\,\AA$^3$, respectively, whereas the experimental value is 486.6\,\AA$^3$.

\paragraph{Hydration layer (HL).}
Several tools model the hydration layer (HL) as a layer with a predefined thickness and a uniform excess density (Fig.~\ref{explsolvent}C). For instance, CRYSOL constructs a homogeneous \SI{3}{\angstrom}-wide layer described by a two-dimensional angular function leading to a simplistic representation of the HL, which may not be suitable for molecules with cavities or non-globular shapes \citep{svergun_crysol_1995}. CRYSOL3 aims to overcome this limitation by incrusting the surface \citep{franke_atsas_2017}) similar to the grid-based density employed by PepsiSAXS \citep{grudinin_pepsi-saxs:_2017}. In contrast, FoXS models the HL by increasing atomic form factors of solvent-exposed atoms \citep{schneidman-duhovny_foxs:_2010}.

The density of HLs of biomolecules likely depends on the type of the water--surface interactions and, therefore, on the chemical composition of the surface. For instance, \citet{kim_saxs/sans_2016} found that acidic amino acids (Glu/Asp) exhibit a tighter HL as compared to basic amino acids (Arg/Lys), an effect that our group recently observed in MD simulations (unpublished data). Likewise, it seems plausible that the HL of maltoside or sodium dodecyl sulfate (SDS) head groups of detergent micelles exhibit different densities as compared to the HL of charged or hydrophobic amino acids. Such chemical specificity of the HL is not captured by implicit solvent methods, explaining why they require fitting of the HL against experimental data.

\paragraph{Risk of overfitting solvent-related parameters}
Owing to the simplified description of the HL and excluded solvent,
implicit-solvent methods require two or three parameters that are
adjusted upon fitting the calculated curve to the experimental curve.
One parameter adjusts the excess density of the HL, while another parameter
adjusts the volumes of the dummy beads that represent the excluded solvent. Some methods
adjust in addition the overall displaced volume. The parameters are typically fitted
by minimizing the residuals between the calculated and experimental SAS curves.
Whether the fitted parameters reflect the physical situation such as the true HL density, or whether they merely absorb errors in the structural model or in the experimental data, is typically not clear. Hence, fitting solvent-related parameter may hide structurally relevant information.

\begin{figure}[h]
	\ifthenelse{\boolean{bShowFigs}}{
		\includegraphics[width=11cm]{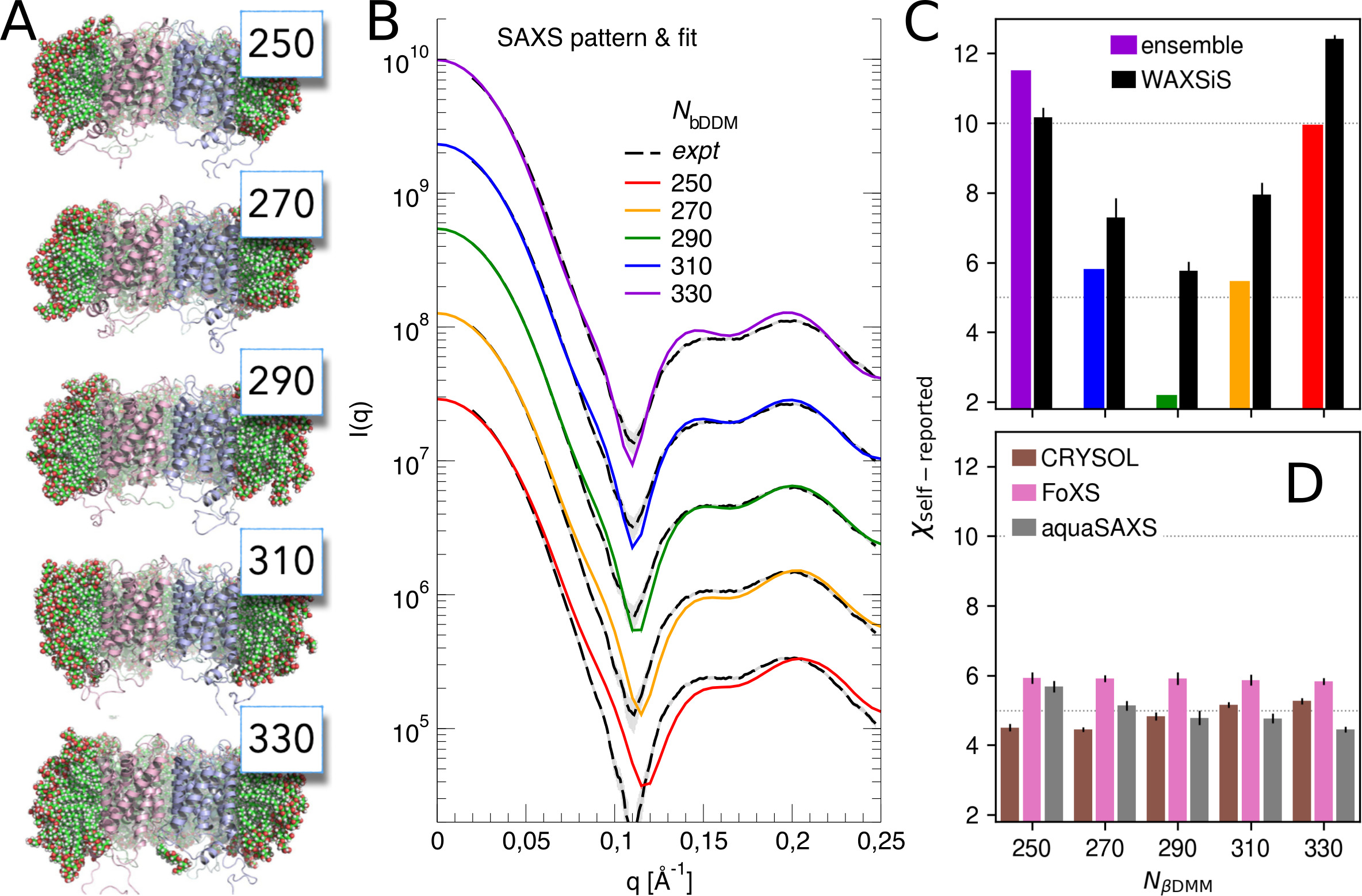}}{}
	\caption{SAXS curve prediction and snapshots of a protein-detergent complex.
		(A) MD simulation systems of aquaporin-0 solubilized in 250 to 330 $\beta$DDM
		detergent molecules simulated with the Charmm36 force field. (B) Explicit-solvent SAXS curve predictions for different number of $\beta$DDM using an aggregated ensemble of \SIrange{90}{100}{\nano\second} MD trajectories (colored lines).
		Experimental data (black dashed line) from \citet{berthaud_modeling_2012}.
		(C) $\chi$-agreement between experimental SAXS curves and explicit-solvent SAXS calculations from free MD simulations (colored bars) or from the WAXSiS webserver (black bars). WAXSiS restrains the DDM molecules in a short simulation, rationalizing the increased $\chi$-values. $N_{\beta\mathrm{DDM}} = 290$ reveals the best agreement with experiment. (D) $\chi$-agreement  by CRYSOL, FoXS, and AquaSAXS, revealing nearly constant $\chi$ due to the adjustment of free parameters.
		Reprinted with permission from \citet{chen_structural_2015}, Copyright 2020 American Chemical Society. \label{protdet}}
\end{figure}

A possible consequence of fitting parameters is illustrated in Fig.\ \ref{protdet}, which presents the validation of MD simulations
of a protein-detergent complex against SAXS data \citep{chen_structural_2015,berthaud_modeling_2012}. The complexes were composed of an aquaporin-0 tetramer solvated in 250 to 330 $\beta$DDM detergent molecules. Computing SAXS curves from the simulations with implicit-solvent methods and fitting the curves to experimental data leads to $\chi$-values that hardly depend on the number of $\beta$DDM molecules, suggesting that alterations in the structural model have been absorbed into the fitting parameters (Fig.\ \ref{protdet}D). Consequently, the experimental number of $\beta$DDM molecules cannot be obtained by comparison with the data. This example illustrates that structurally relevant information may be hidden upon fitting free parameters. In contrast, when using explicit-solvent SAXS calculations described below, the $\chi$-agreement of models with different $\beta$DDM numbers with the data can be distinguished because no solvent-related parameters are fitted, thereby revealing that the model with 290 $\beta$DDM exhibits the best agreement with the data (Fig.\ \ref{protdet}C).



\section{Explicit-solvent SAS predictions with the WAXSiS method}

Several methods for predicting SAS curves have been presented that model both the excluded solvent and the HL with explicit-solvent MD simulations \citep{oroguchi_intrinsic_2009,park_simulated_2009,kofinger_atomic-resolution_2013,chen_validating_2014}. Here, we focus on the WAXSiS method, which is based on the formalism by \citet{park_simulated_2009}. However, whereas the implementation by Park \textit{et al.} required a constrained biomolecule, the WAXSiS method allows SAS predictions from MD simulations of flexible biomolecules by defining the SAS-contributing solvent with a spatial envelope. In addition, the WAXSiS method corrects (i) imprecise densities of water models and (ii) small density mismatches between the biomolecule and the pure-solvent simulation systems, both of which may influence the SAS prediction significantly. A careful comparison of different implicit and explicit SAXS prediction methods has been presented by \citet{bernetti_comparing_2021}.

\begin{figure}[h]
	\ifthenelse{\boolean{bShowFigs}}{
		\includegraphics[width=12cm]{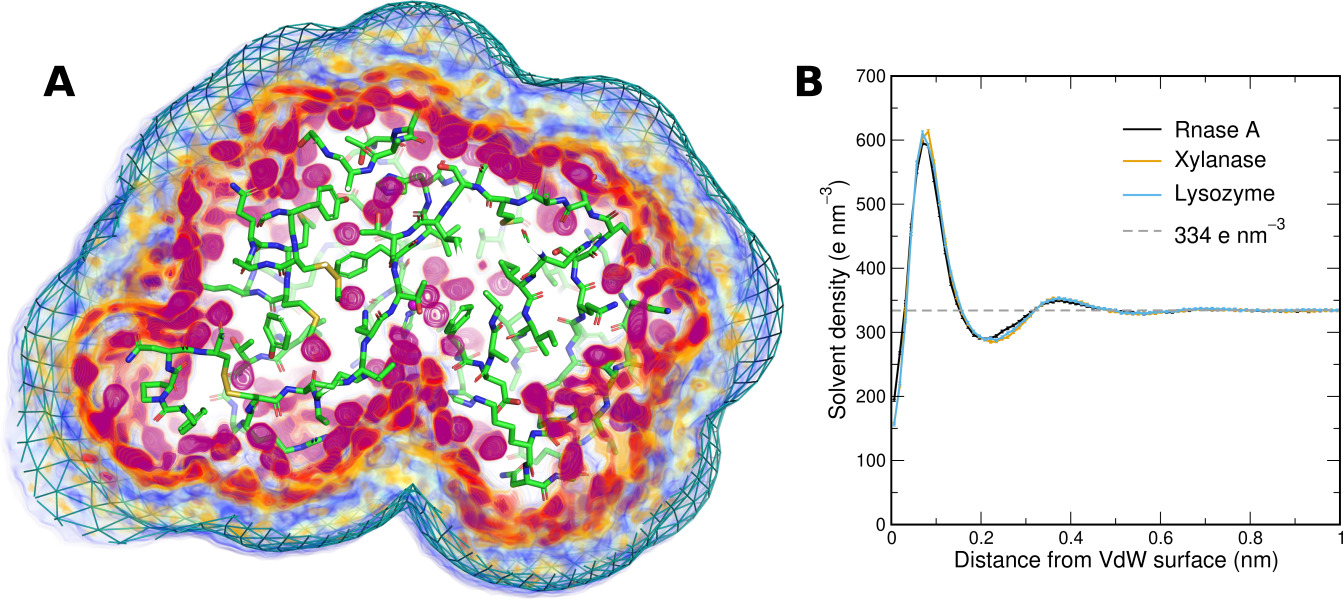}}{}
	\caption{(A) Solvent electron density around Rnase A, taken from an explicit-solvent MD simulations with position restraints on all heavy atoms, revealing the marked structure of the hydration layer. Only solvent density inside an envelope (blue mesh) is shown, where the envelope was constructed at a 7\AA{} distance from the protein surface. (B) Solvent density as function of the distance from the Van der Waals surface for three different proteins (see legend), averaged over the protein surface. Simulations were carried out with the Amber99sb-ildn/Tip3p force fields. Here, the solvent density was scaled with a small constant factor to match the experimental bulk density of 334\,e\,nm$^{-3}$, as implemented in GROMACS-SWAXS to correct imprecise densities of some water models. \label{HL}}
\end{figure}

In the WAXSiS method, a spatial envelope is defined that encloses the biomolecule and the heterogeneous density of the HL (Fig.~\ref{explsolvent}A, Fig.~\ref{HL}A). Hence, the HL structure is fully defined by the force field and does not require a free fitting parameter. The buffer subtraction is carried out by computing the scattering intensity difference between (a) the biomolecule with the envelope-enclosed solvent and (b) an explicit-solvent pure-buffer simulation, from which an identical volume is taken with the help of the same envelope (Fig~\ref{explsolvent}B). Thereby, the displaced solvent is modeled with atomic detail and without need of knowing atomic volumes of solute atoms.

Compared to implicit-solvent methods, the explicit-solvent calculations are computationally far more expensive as they require MD simulations, but they have several crucial advantages:
\begin{itemize}
\item[(i)] The calculations do not require any free solvent-related parameters that must be fitted to experimental data. Thus, the amount of structural information, which can be extracted from the low-information SAS data, is not further reduced. By circumventing the fitting of the HL density, the radius of gyration is not adjusted against the data.
\item[(ii)] The use of ``reduced form factors'' is avoided and, with this, the uncertainties subject to atomic volumes. This may explain the better accuracy compared to implicit-solvent methods for predicting the SAS intensities of heterogeneous systems (such as protein-detergent complexes) and for molecules with non-globular geometries \citep{ivanovic_temperature-dependent_2018,ivanovic_small-angle_2020}.
\item[(iii)] The explicit-solvent description remains valid at wide scattering angles ($q>$\SI{1.5}{\angstrom}), where the internal  structure of water becomes relevant \citep{park_simulated_2009}.
\item[(iv)] By computing the SAS patterns from an MD trajectory, the method naturally accounts for thermal fluctuations, which influence the SAS patterns at remarkably small scattering angles of $q\gtrsim 0.25$\,\AA$^{-1}$ \citep{tiede_protein_2002,moore_effects_2014,chen_validating_2014,chen_structural_2015}.
\end{itemize}

\begin{figure}[h]
	\ifthenelse{\boolean{bShowFigs}}{
		\includegraphics[width=12cm]{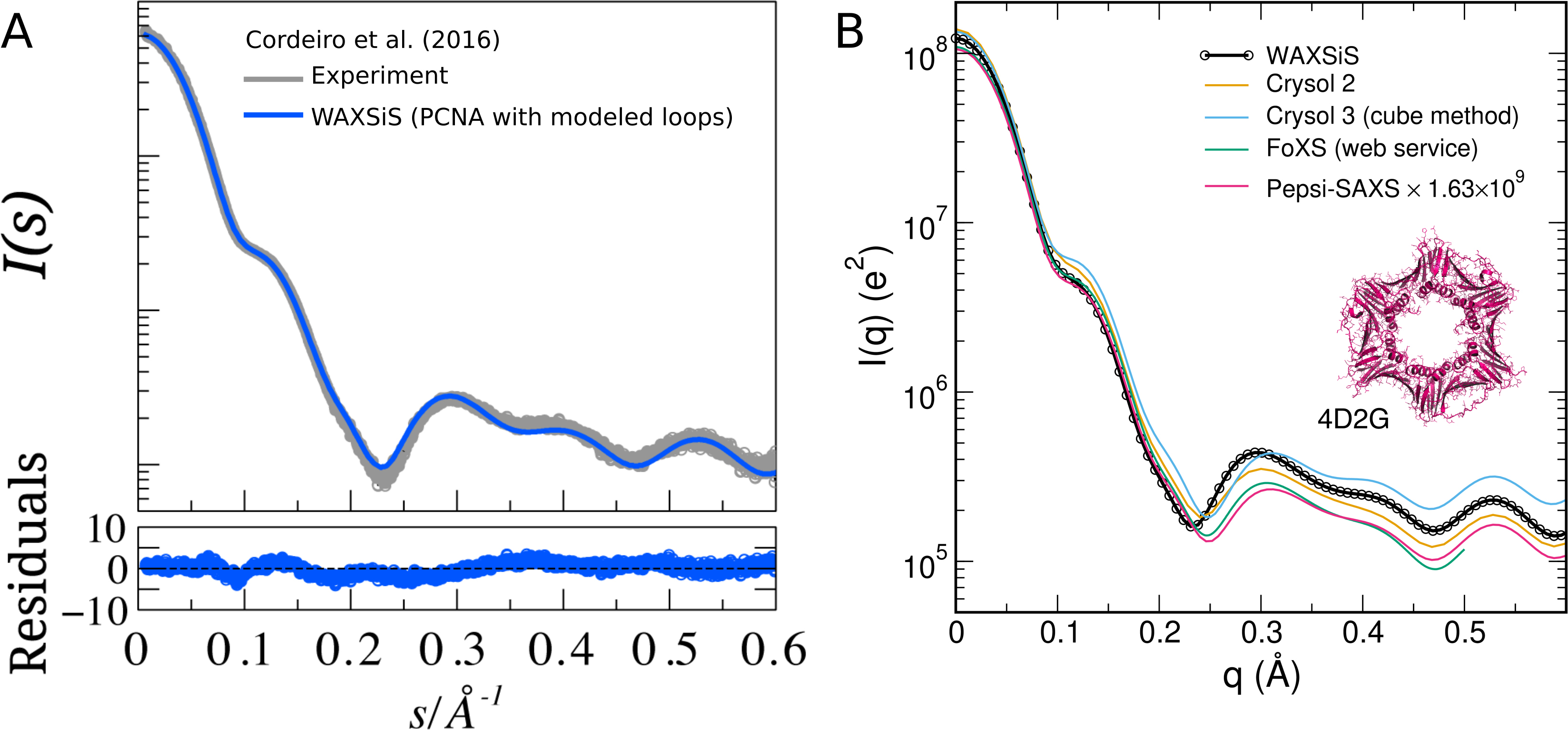}}{}
	\caption{(A) Experimental SAXS data of the ring-shaped protein PCNA (grey) and SAXS curve computed with the WAXSiS model from a PCNA model with modeled loops $\left(\chi^2=1.35\right)$ \citep{cordeiro_disentangling_2017}. (B) Comparison of SAXS curves of the PCNA crystal structure (pdb code 4D2G) computed with different methods with default settings (see legend). Panel A adapted and reused with permission from \citet{cordeiro_disentangling_2017}.  \label{methods}}
\end{figure}

As an example, Fig.~\ref{methods}A compares the experimental SAXS curve of the ring-shaped protein PCNA with a curve computed with the WAXSiS method. The WAXSiS curve was computed from the 4D2G crystal structure of PCNA, to which missing termini were added \citep{cordeiro_disentangling_2017}, and the computed curve was henceforth scaled with a constant factor ($I_\mathrm{scaled}(q) = f\,I_\mathrm{WAXSiS}(q)$). No other parameters were adjusted.  Figure \ref{methods}B compares SAXS curves computed from the 4D2G structure (without termini) computed with WAXSiS and with several implicit-solvent methods. All methods used default settings, and the calculated curves were not fitted against the experiment. Evidently, the predicted SAXS curves differ significantly, presumably owing to different modeling of the hydration layer and excluded solvent.

\section{Theory}

The experimental SAS intensity is given by the difference between the scattering intensity of the solution $I_\mathrm{sample}(q)$ and the solvent $I_\mathrm{solvent}(q)$:
\begin{equation}
 I(q) = I_\mathrm{sample}(q) - I_\mathrm{solvent}(q)
\end{equation}
To calculate this excess intensity $I(q)$ from MD trajectories, the WAXSiS method follows the formalism by \citet{park_simulated_2009}.
The low-dilution limit is considered, such that correlations between different solute molecules can be neglected and the scattering experiment can be modeled by a single solute molecule in a solvent bath. A spatial envelope is constructed around the solute  including the hydration layer (Fig.~\ref{explsolvent}A). The instantaneous electron density of the solute system $A(\textbf{r})$ and of the solvent system $B(\textbf{r})$ is divided into the electron density inside and outside of the envelope, indicated by subscripts $\mathrm{i}$ and $\mathrm{o}$, respectively,
\begin{align}
 A(\textbf{r}) &= A_\mathrm{i}(\textbf{r}) + A_\mathrm{o}(\textbf{r}) \, , \label{electrondensities1}\\
 B(\textbf{r}) &= B_\mathrm{i}(\textbf{r}) + B_\mathrm{o}(\textbf{r}) \, .
 \label{electrondensities2}
\end{align}
Assuming that the envelope is large enough such that density correlations between the inside and the outside of the envelope are due to bulk water, the excess scattering intensity only requires knowledge of the Fourier transforms of the electron densities \textit{inside} the envelope $\tilde{A}_\mathrm{i}(\textbf{q})$ and $\tilde{B}_\mathrm{i}(\textbf{q})$:
\begin{equation}
I(q) = \big\langle D(\textbf{q}) \big\rangle_\Omega \, ,
\end{equation}
where
\begin{equation}
   D(\textbf{q})  = \bigl\langle \bigl|\tilde{A}_\mathrm{i}(\textbf{q})\bigr|^{2} \bigr\rangle ^{(\omega)} - \bigl\langle \bigl|\tilde{B}_\mathrm{i}(\textbf{q})\bigr|^{2} \bigr\rangle ^{(\omega)} + 2 \mathrm{Re}\bigl[ -\bigl\langle\tilde{B}_\mathrm{i}^*(\textbf{q})\bigr\rangle^{(\omega)}\bigl\langle \tilde{A}_\mathrm{i}(\textbf{q}) - \tilde{B}_\mathrm{i}(\textbf{q}) \bigr\rangle^{(\omega)}].
\end{equation}
Here, $\langle...\rangle_\Omega$ denotes the orientational average in $\mathbf{q}$-space,  $\langle...\rangle^{(\omega)}$ the average over solute and solvent fluctuations at fixed solute orientation $\omega$, and $\mathrm{Re}$ is the real part. In practice, $\langle...\rangle^{(\omega)}$ represents the average over MD frames after superimposing the solute onto a reference structure of orientation $\omega$. $\tilde{A}_\mathrm{i}(\textbf{q})$ and $\tilde{B}_\mathrm{i}(\textbf{q})$ are calculated using the atomic form factors $f_j(q)$ and coordinates $\textbf{r}_j$ of atom $j$,
\begin{equation}
 \tilde{A}_\mathrm{i}(\textbf{q}) = \sum_{j=1}^{N_A} f_j(q) e^{-i\textbf{q}\cdot\textbf{r}_j} \, ,
\end{equation}
where $N_A$ is the number of atoms within the envelope in the respective MD frame (Fig.~\ref{explsolvent}A).  $\tilde{B}_i(\textbf{q})$ is calculated analogously over the solvent atoms within the envelope (Fig.~\ref{explsolvent}B).
The form factors are approximated using the Cromer-Mann parameters $a_k,b_k$ and $c$ of atom $j$ \citep{cromer_x-ray_1968},
\begin{equation}
 f_j(q) = \sum_{k=1}^4 a_k e^{-b_k(q/4\pi)^2} + c \, .
\end{equation}
The formalism is identical for SANS calculations, except that the atomic form factors $f_j(q)$ are replaced by the coherent neutron scattering lengths $b_j$ of the scattering atoms \citep{chen_combined_2019,dias-mirandela_merging_2018}.

\subsubsection*{Comparing to experimental data}
To compare a predicted curve with experimental data, the experimental curve $I_\mathrm{exp}$ can be fitted to the calculated curve $I_\mathrm{c}$ by minimizing
\begin{equation}
 \chi^2 = N_q^{-1} \sum_{i=1}^{N_q}\frac{\bigl[ I_\mathrm{c}(q_i) - \bigl(f I_\mathrm{exp}(q_i)+c\bigr)\bigr]^2}{\sigma_i^2} \, ,
\end{equation}
where $N_q$ denotes the number of $q$-points, and $\sigma_i$ are the experimental errors. Besides the overall scale $f$, an offset $c$ may be fitted to account for small experimental uncertainties from the buffer subtraction. Here, the experimental curves are fitted instead of the calculated curves, because the calculated curves are predicted without any free parameters. In practice, it is advisable to fit also without the offset $c$ and, thereby, to test whether $c$ absorbs only a small constant offset, as desired, or whether fitting the offset would hide a structurally relevant discrepancy between $I_\mathrm{exp}$ and $I_\mathrm{c}$.

\section{Workflow}

\begin{figure}[h]
	\ifthenelse{\boolean{bShowFigs}}{
		\includegraphics[width=12cm]{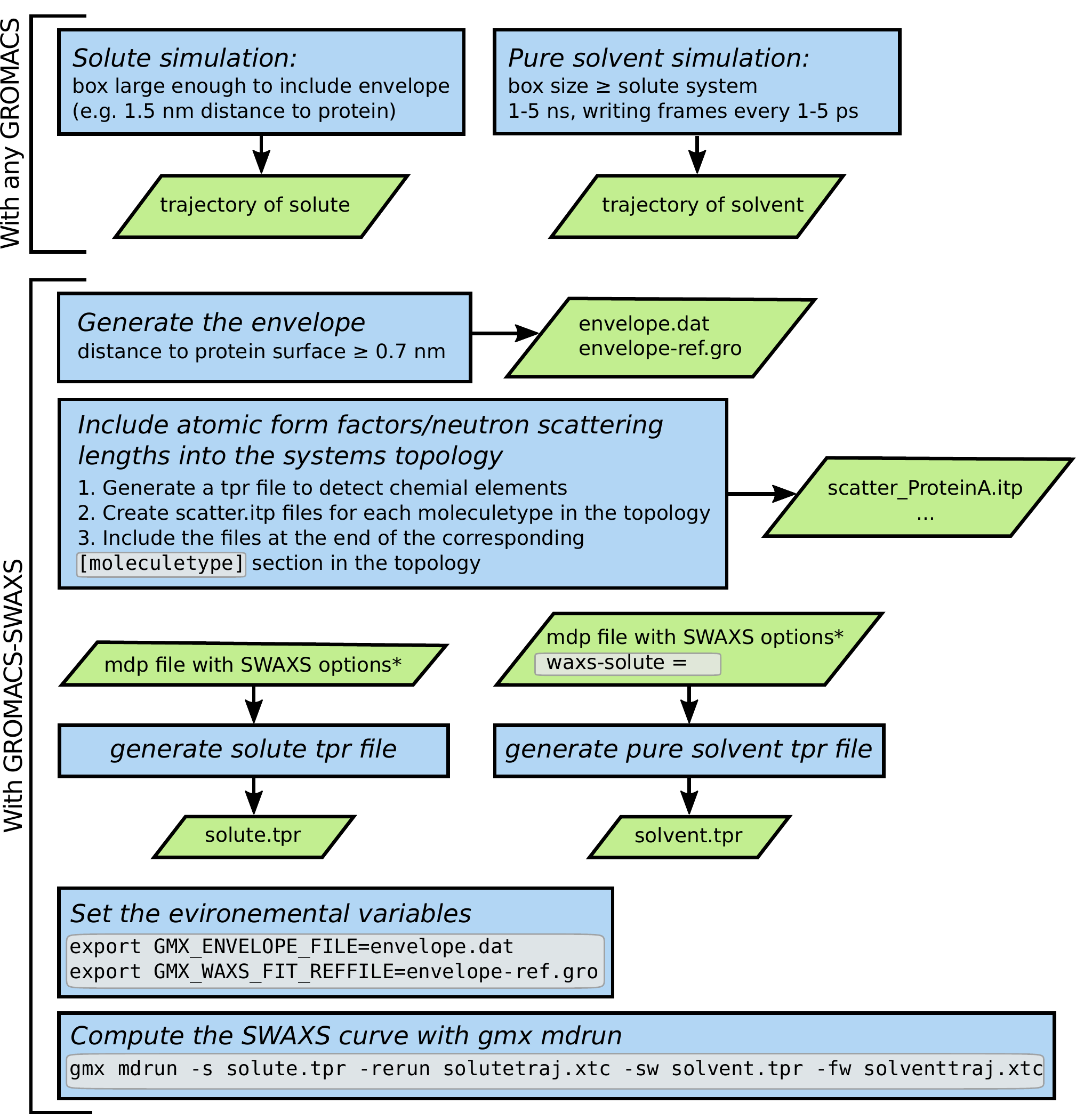}}{}
	\caption{Workflow for SAS curve predictions with GROMACS-SWAXS. *Suggested mdp parameters are shown in listing \ref{mdpparam}. \label{workflow}}
\end{figure}

We present a brief workflow for computing a SAS curve from an explicit-solvent MD simulation (Fig.~\ref{workflow}). Because the SAS calculations are implemented into an extension of GROMACS, it is convenient (but not required) to run the MD simulation with GROMACS as well. This workflow is automated also on the web server WAXSiS (https://waxsis.uni-saarland.de). However, the MD simulations on WAXSiS are carried out with the Yasara software which allows fully automated MD simulations even in presence of ligands or modified amino acids \citep{krieger_new_2015}. Several additional tutorials and more detailed documentation for GROMACS-SWAXS are available at https://cbjh.gitlab.io/gromacs-swaxs-docs.

\subsection{A: With any GROMACS}

\noindent
1. Run an MD simulation of your biomolecule in solution with either GROMACS-SWAXS or any other GROMACS version. Choose the simulation box large enough, such that the envelope (see below) will fit into the box. To generate the box use, e.g.,
\begin{lstlisting}
 gmx editconf -d 1.5 -f protein.gro -bt dodecahedron -o box.gro
\end{lstlisting}
where for the \colorbox{backcolour}{\lsin{-d}} option a distance between \colorbox{backcolour}{\lsin{1.5}} and \colorbox{backcolour}{\lsin{2.0}} is usually sufficient.

\noindent
2. Run an MD simulation of pure solvent for the buffer subtraction. Use the same water model and MD parameter (mdp) file as used for the solute simulations to ensure that solute and solvent simulations exhibit identical bulk solvent densities, as required for an accurate buffer subtraction. The solvent simulation box should be at least as large as your solute simulation box. Produce \SIrange{1}{5}{\nano\second} of simulation, writing the simulation frames every \SIrange{1}{5}{\pico\second}, i.e. $\sim$1000 frames.

\subsection{B: With GROMACS-SWAXS}
When the simulations of solute and pure solvent have finished, we continue with preparing and running the SAS curve calculation. Once the steps below are understood, they can be scripted to automate the SAS curve calculation.

\noindent
1. Download and compile GROMACS-SWAXS (https://gitlab.com/cbjh/gromacs-swaxs), and load it into your path:
\begin{lstlisting}
source /path/to/gromacs-swaxs/bin/GMXRC
\end{lstlisting}

\subsubsection{Specify the atomic form factors/neutron scattering lengths}
\noindent
2.  Generate a run-input (tpr) file using \colorbox{backcolour}{\lsin{gmx grompp}} of GROMACS-SWAXS to detect the chemical elements, which are used below for assigning the Cromer-Mann parameters or neutron scattering lengths. Chemical elements are recognized using  atom names and masses to ensure
that a C$_\alpha$ carbon atom (C) is not confused with calcium (Ca), Fluor (F) not with Iron (Fe), etc. Atomic masses are reliably provided by a tpr file. An empty mdp file may be used for this step:
\begin{lstlisting}
rm -f anymdp.mdp; touch anymdp.mdp
gmx grompp -f anymdp.mdp -p topol.top -o tmp.tpr
\end{lstlisting}

\noindent
3. Now, the previously created tpr file is used to generate a \colorbox{backcolour}{\lsin{scatter.itp}} file for each molecule, which contains the Cromer-Mann parameters or the neutron scattering lengths. If the solute contains non-default groups, such as chromophore, a heme group, or similar, first prepare an index file with the \textit{entire} solute. Then run:
\begin{lstlisting}
gmx genscatt -s tmp.tpr [-n index.ndx] [-vsites]
\end{lstlisting}
Select your \textit{complete} solute, such as Protein, Protein\_HEME or Protein\_Chromophore. If you use virtual sites, use \colorbox{backcolour}{\lsin{-vistes}} and select Prot-Masses (or a group with heme, the chromophore, etc.). The \colorbox{backcolour}{\lsin{-visites}} option instructs \colorbox{backcolour}{\lsin{gmx genscatt}} that atomic masses deviate from the physical masses, as common when using virtual sites.  \colorbox{backcolour}{\lsin{gmx genscatt}} writes one itp file for each molecule type. These must be added to each moleculetype definition in the topology, similar to the \colorbox{backcolour}{\lsin{#include "posres.itp"}} line for defining position restraints.

\subsubsection{Generating the envelope}
\label{envtext}

\noindent
4. Generate the envelope from the trajectory of the solute simulations
\begin{lstlisting}
gmx genscatt -s tmp.tpr -f solutetraj.xtc -d 0.7
\end{lstlisting}
The output \colorbox{backcolour}{\lsin{envelope.dat}} lists inner and outer radii of the envelope surface, whereas \colorbox{backcolour}{\lsin{envelope-ref.gro}} is a reference structure with orientation $\omega$ used to superimpose the solute frames onto the envelope (see Theory). The envelope may be visualized with PyMol with the file \colorbox{backcolour}{\lsin{envelope.py}}.

\subsubsection{Generating the solute and solvent run-input files}

\noindent
5. Generate tpr files with \colorbox{backcolour}{\lsin{gmx grompp}} and with an mdp file that contains all SAS-specific parameters, as presented in Listing~\ref{mdpparam}.
\begin{lstlisting}
gmx grompp -f rerun.mdp -c solutesystem.gro [-n index.ndx] -o solute.tpr
\end{lstlisting}
In addition, a tpr file of the pure-solvent system is required, prepared with a mdp file with an empty solute entry (\colorbox{backcolour}{\lsin{waxs-solute =}})
\begin{lstlisting}
gmx grompp -f solvent.mdp -c solventsystem.gro -o solvent.tpr
\end{lstlisting}

\begin{lstlisting}[caption={A typical set of SAS-specific mdp parameters for calculating a SWAXS/SANS curve with GROMACS-SWAXS. The calculation of SANS curves is optional, as indicated by square brackets.},label=mdpparam,float]
; read scattering info from topology
define = -DSCATTER
; turn on SAXS calc. and, optionally, multiple SANS calc.
scatt-coupl = xray [neutron neutron ...]
; solute group
waxs-solute   = Protein ; or Protein-Masses
; solvent group
waxs-solvent  = Water_and_ions
; rotational fit group as used with gmx genenv
waxs-rotfit   = C-alpha
; or use a PBC atom near the geometric center, as suggested by gmx genenv
waxs-pbcatom  = -1
; for uniform average over all simulation frames
waxs-tau      = 0
; number of q-points
waxs-nq       = 101
; qmin and qmax in nm^(-1)
waxs-startq   = 0
waxs-endq     = 10
; number of q-vectors for spherical average, use ~0.2*(D*qmax)^2
waxs-nsphere  = 1500
; experimental solvent density in e/nm3, for solvent density correction
waxs-solvdens = 334
; Use I(sample)-I(buffer) as buffer subtraction scheme
waxs-correct-buffer = no
; D2O concentrations of SANS calculations
waxs-deuter-conc = [0.42 1]
\end{lstlisting}

\subsubsection{Compute the SAS curve}

\noindent
6. Finally, compute the SAS curves from the solute simulations with the rerun functionality of \colorbox{backcolour}{\lsin{gmx mdrun}}. The envelope files are specified with environment variables, as follows:
\begin{lstlisting}
export GMX_ENVELOPE_FILE=envelope.dat
export GMX_WAXS_FIT_REFFILE=envelope-ref.gro
gmx mdrun -s solute.tpr -rerun solutetraj.xtc \
          -sw solvent.tpr -fw solventtraj.xtc
\end{lstlisting}
The SAS calculations strongly benefit from the use of a GPU or from using multiple CPU cores, for instance using \colorbox{backcolour}{\lsin{gmx mdrun -ntomp 16 -gpu_id 0 ...}}. The SAS curves are written to the files \colorbox{backcolour}{\lsin{waxs_final.xvg}}.

\section{Practical considerations}

\subsection{Convergence}

The number of simulation frames required to receive a converged SAS curve depends on the contrast between solute and solvent, where a lower contrast leads to slower convergence.
The contrast is lower if the hydration layer is large compared to the biomolecule, i.e., if the envelope contains mostly water. Specifically, small biomolecules or intrinsically disordered proteins (IDPs) require a larger number of simulation frames as compared to larger or globular biomolecules. As a reference, for a smaller protein such as the GB3 domain, approximately 300 frames are required to obtain a well-converged SAS curve. A larger protein such as glucose isomerase requires approx.\ 70 frames. The statistical errors listed in \colorbox{backcolour}{\lsin{waxs_final.xvg}} are computed with error propagation, assuming that each simulation frame provides an independent  solvent structure, which is fulfilled when using frames with time spacing $\gtrsim 1$\,ps.

For IDPs, the convergence should be carefully assessed, for instance by dividing the trajectory of \textit{both} the biomolecule and the buffer into ten independent blocks, and then comparing the ten SAS curves computed from the blocks. The standard error over the blocks provides a rigorous measure for statistical uncertainty.

\subsection{Orientational average}

The parameter $J$ (mdp option waxs-nsphere) determines the number of $\textbf{q}$-vectors used to take the orientational average. It should be taken as $J \approx \alpha(Dq_\mathrm{max})^2$. Where $D$ is the maximum diameter of the solute, $q_\mathrm{max}$ the maximum momentum transfer (mdp option waxs-endq), and the constant $\alpha$ determines the accuracy of the orientational average.
For most biomolecules, $\alpha=0.2$ is sufficient, while for excessively elongated rod-like structures a value of  $\alpha=0.5$ may be required.

\subsection{Envelope size}

The distance of the envelope from the biomolecular surface (specified with {\tt gmx genenv -d}) should be at least \SIrange{0.6}{0.7}{\nano\meter}, thereby containing the density modulations along the hydration layers.
For highly charged biomolecules, a larger envelope may be used to account for the entire counter ion cloud. We found previously that a distance of $\sim$3 times the Debye length is required to include most effects from the counter ion cloud on the radius of gyration of glucose isomerase \citep{ivanovic_quantifying_2018}. At a salt concentration of 100\,mM, this may require an envelope distance of $\sim$3\,nm and, hence, large simulation systems.

\subsection{Solvent density correction}
A solvent density correction is necessary for two reasons:
(i) Due to finite-size effects, the bulk solvent densities and correlations may slightly differ between the solute simulation and the pure-solvent simulation \citep{kofinger_atomic-resolution_2013}.
(ii) The density of certain water models, such as the popular Tip3p model, deviate from the experimental water densities. The parameter for solvent density correction should be chosen to match the experimental electron density of the solvent (mdp option \colorbox{backcolour}{\lsin{waxs-solvdens}}).

\subsection{Atomic fluctuations}
Structural fluctuations influence the SAS profile. Depending on the scientific question,
position restraints may or may not be applied to the biomolecule, for instance with the aim
to obtain the SAS curve for structure with given backbone coordinates or for a flexible biomolecular ensemble, respectively.
To resolve the structure of the HL, it may be useful to restrain all heavy atoms of the biomolecule to avoid that side chain fluctuations smear out the HL structure, as used to obtain the solvent densities shown in Fig.~\ref{HL}.

\subsection{Water models}

We validated previously that most popular water force fields lead to nearly identical SAS curves at moderate scattering angles, suggesting that most water models exhibit similar packing at the biomolecular surface \citep{chen_validating_2014}. Only at very wide angle of $q\gtrsim 2$\,\AA$^{-1}$, where the water scattering is dominant, large effects from the water models are visible. However, we recently observed that water models with increased dispersion interactions such as the Tip4p2005s or Tip4p-D may impose tighter packing on the protein surface, leading to slightly increased radii of gyration (unpublished data) \citep{best_balanced_2014,piana_water_2015}. These observations open a novel route for validating protein--water interactions by modern force fields against experimental SAS data.

\subsection{Computational costs}

Run times of SAS predictions depend on the number of atoms, on the applied convergence criteria, and on the hardware. Typical costs may be illustrated by the execution time for different proteins on the WAXSiS webserver at https://waxsis.uni-saarland.de, which is currently (by August 2022) equipped with a 16-core AMD Ryzen 9 processor and an Nvidia RTX 3070Ti graphics card. 
Using the convergence setting ``normal'', which is suitable for tests, SAS predictions of lysozyme (PDB id 1LYS), glucose isomerase (PDB id 1MNZ), and RNA polymerase II (PDB id 1TWF) required \SI{2.7}{min}, \SI{4.8}{min}, and \SI{15.2}{min}, respectively. Using the convergence setting ``thorough'', which is recommended for publications, the execution times for the same biomolecules increased to \SI{10.8}{min}, \SI{14.2}{min}, and \SI{36.3}{min}, respectively. These numbers may change as the hardware improves or as the convergence criteria are adapted in future releases of the WAXSiS webserver.

\section{Summary}

Both implicit- and explicit-solvent SAS prediction methods are highly valuable for the structural interpretation of experimental SAS methods. Implicit-solvent methods are computationally efficient, thus allowing high-throughput calculations on a laptop. They are unequivocally capable of validating the overall shape of structural models against SAS data, and to reveal effects from larger conformational transitions. To harvest structural details, as encoded in high-precision SAS data collected from modern SEC-SAXS or SEC-SANS experiments, explicit-solvent methods are needed. They are (i) based on an accurate representation of the hydration layer and the excluded solvent, thereby avoiding solvent-related fitting parameters, (ii) are accurate for inhomogeneous systems since they do not require tabulated atomic volumes, and (iii) include effects from structural fluctuations on SAS curves. The WAXSiS method presented here may be further used for structure refinement simulations by restraining an MD simulation to experimental SAS data \citep{chen_interpretation_2015, chen_combined_2019}. More details on structure refinement are provided in a chapter of Part B of this monograph.

Running a SAS prediction on the WAXSiS webserver requires between 3 and 40\,min including the MD simulation.
On a fast workstation or a computer cluster, tens to hundreds of explicit-solvent calculations are feasible within days, suggesting that the cost for explicit-solvent SAS calculations are negligible as compared to the costs for experimental sample preparation or for data collection at the beamline. Therefore, we expect that the methods presented here will become a routinely used tool for SAS-based structural biology.


\section{Acknowledgements}
We thank Tobias Fischbach for support with preparing Figure 2A. This study was supported by the Deutsche Forschungsgemeinschaft (HU 1971-3/1).

\bibliographystyle{apa}
\bibliography{sas.bib,Zotero}


\end{document}